\documentstyle[12pt]{article}

\begin{document}

\title{\bf {Correspondence between Holographic and
Gauss-Bonnet dark energy models}}
\author{\normalsize{M. R. Setare$^{1}$\thanks{%
E-mail: rezakord@ipm.ir}  \, and \,E.~N.~Saridakis $^{2}$\thanks{%
E-mail: msaridak@phys.uoa.gr} }\\
\newline
\\
{\normalsize \it $^1$ Department of Science, Payame Noor
University, Bijar, Iran}
\\
{\normalsize \it $^2$ Department of Physics, University of Athens,
GR-15771 Athens, Greece}
\\
}
\date{\small{}}

\maketitle
\begin{abstract}

In the present work we investigate the cosmological implications
of holographic dark energy density in the Gauss-Bonnet framework.
By formulating independently the two cosmological scenarios, and
by enforcing their simultaneous validity, we show that there is a
correspondence between the holographic dark energy scenario in
flat universe and the phantom dark energy model in the framework
of Gauss-Bonnet theory with a potential. This correspondence leads
consistently to an accelerating universe. However, in general one
has not full freedom of constructing independently the two
cosmological scenarios. Specific constraints must be imposed on
the coupling with gravity and on the potential.

 \end{abstract}

\newpage

\section{Introduction}

Nowadays it is strongly believed that the universe experiences an
accelerated expansion. Recent observations from type Ia supernovae
\cite{SN} in associated with Large Scale Structure \cite{LSS} and
Cosmic Microwave Background anisotropies \cite{CMB} have provided
main evidence for this cosmic acceleration. In order to explain
this interesting behavior, many theories have been proposed.
Although it is widely accepted that the cause which drives the
acceleration is the so called dark energy, its nature and
cosmological origin still remain enigmatic at present. One recent
proposal is the dynamical dark energy scenario (see
\cite{DEreview} and references therein), since the cosmological
constant puzzles may be better interpreted by assuming that the
vacuum energy is cancelled to exactly zero by some unknown
mechanism and introducing a dark energy component with a
dynamically variable equation of state. The dynamical dark energy
paradigm is often realized by some scalar field mechanism which
suggests that the energy form with negative pressure is provided
by a scalar field evolving under a suitable potential.

In addition, many string theorists have devoted to understand and
shed light on the cosmological constant or dark energy within the
string framework. The famous Kachru-Kallosh-Linde-Trivedi (KKLT)
model \cite{kklt} is a typical example, which tries to construct
metastable de Sitter vacua in the light of type IIB string theory.
Furthermore, string landscape idea \cite{landscape} has been
proposed for shedding light on the cosmological constant problem
based upon the anthropic principle and multiverse speculation.
Although we are lacking a quantum gravity theory today, we still
can make some attempts to probe the nature of dark energy
according to some principles of quantum gravity. Currently, an
interesting attempt in this direction is the so-called
``holographic dark energy'' proposal
\cite{Cohen:1998zx,Hsu:2004ri,Li:2004rb}. Such a paradigm has been
constructed in the light of the holographic principle of quantum
gravity theory, and thus it presents some interesting features of
an underlying theory of dark energy \cite{holoprin}. Furthermore,
it may simultaneously provide a solution to the coincidence
problem, i.e why matter and dark energy densities are comparable
today although they obey completely different equations of motion
\cite{Li:2004rb}. The holographic dark energy model has been
extended to include the spatial curvature contribution
\cite{nonflat} and it has also been generalized in the braneworld
framework \cite{bulkhol}. Lastly, it has been tested and
constrained by various astronomical observations
\cite{obs3,HG,observHDExray,observHDECMB,observHDE,cmb3}.

Since holographic energy density corresponds to a dynamical
cosmological constant, we need a dynamical framework, instead of
general relativity, to consistently accommodate it. Therefore, it
is interesting to investigate it under the Brans-Dicke theory
\cite{{gong},{mu}, {tor},{set1}}. As it is known, Einstein's
theory of gravity may not describe it correctly at very high
energy. The simplest alternative to general relativity is
Brans-Dicke scalar-tensor theory \cite{bd}, and amongst the most
popular modified-gravity attempts, which may successfully describe
the cosmic acceleration, is the $f(R)$-gravity. Very simple
versions of such a theory, like $1/R$ \cite{1} and $1/R + R^2$
\cite{2}, may lead to the effective quintessence/phantom late-time
universe. Another proposal, closely related to the low-energy
string effective action, is the scalar-Gauss-Bonnet gravity
\cite{3}, which can be considered as a form of gravitational dark
energy.

In the present paper we are interested in investigating the
conditions under which we can obtain a correspondence between
holographic and Gauss-Bonnet models of dark energy, i.e to examine
holographic dark energy in a spatially flat Gauss-Bonnet universe.

\section{Gauss-Bonnet Dark Energy}

In this section we formulate a Gauss-Bonnet model for dark energy
\cite{3,4,Nojiri2}. As usual, as a candidate for dark energy we
consider a scalar field $\phi$, which is moreover coupled to
gravity through the higher-derivative (string-originated)
Gauss-Bonnet term. The corresponding action is given by
\begin{equation}
S=\int d^{4}x\,\sqrt{g}\,\left[
\frac{1}{2\kappa^2}R-\frac{\gamma}{2}\partial_{\mu}\phi\partial^{\mu}\phi-V(\phi)+f(\phi)G\right]
, \label{action}
\end{equation}
where $\kappa^2=8\pi G$ and $\gamma=\pm1$. For a canonical scalar
field $\gamma=1$, but we extend the model to $\gamma=-1$ which
corresponds to phantom behavior. In (\ref{action}) $G$ stands for
the Gauss-Bonnet combination which is explicitly given as:
\begin{equation}
G=R^2-4R_{\mu\nu}R^{\mu\nu}+R_{\mu\nu\rho\sigma}R^{\mu\nu\rho\sigma},
\label{GB}
\end{equation}
where $R_{\mu\nu\rho\sigma}$, $R_{\mu\nu}$ are respectively the
Riemann and Ricci tensors and $R$ is the curvature scalar of the
spacetime with metric $g_{\mu\nu}$. Finally, the coupling with
gravity constitutes of a function $f(\phi)$. In the following we
will concentrate on the spatially flat Robertson-Walker universe
with
\begin{equation}\label{met}
ds^{2}=-dt^{2}+a(t)^{2}(dr^{2}+r^{2}d\Omega^{2}),
\end{equation}
thus we impose $k=0$ in (\ref{action}).

The equations of motion can be easily derived from (\ref{action}),
and the result is \cite{4}:
\begin{equation}\label{2}
\frac{\gamma}{2}\dot{\phi}^{2}-V(\phi)+16f'(\phi)\dot{\phi}H\frac{\ddot{a}}{a}+8\left[f'(\phi)\ddot{\phi}
+f''(\phi)\dot{\phi}^{2}\right]H^2=p_\Lambda
\end{equation}
for the  scale factor, and
\begin{equation}\label{3}
\gamma\left[\ddot{\phi}+3H\dot{\phi}+\frac{V'(\phi)}{\gamma}\right]=24f'(\phi)H^2\frac{\ddot{a}}{a}
\end{equation}
for the scalar field. Furthermore, we obtain a constraint
equation, namely:
\begin{equation}\label{1}
\frac{\gamma}{2}\dot{\phi}^{2}+V(\phi)-24f'(\phi)\dot{\phi}H^3=\rho_\Lambda.
\end{equation}
In the expressions above, $p_\Lambda$ and $\rho_\Lambda$ are the
pressure and energy density due to the scalar field and the Gauss
Bonnet interaction \cite{Nojiri2}, which are identified as the
corresponding quantities of dark energy.

\section{Holographic Dark Energy}

Let us describe briefly the holographic dark energy model
\cite{Cohen:1998zx,Hsu:2004ri,Li:2004rb}.
 In this dark-energy-paradigm one determines an appropriate quantity to serve
 as an infrared cut-off for the theory, and imposes the constraint
 that the total vacuum energy in the corresponding maximum volume
 must not be greater than the mass of a black hole of the same
 size. By saturating the inequality one identifies the acquired
 vacuum energy as holographic dark energy. Although the choice of the IR cut-off has raised a discussion in the literature
\cite{Li:2004rb,gong,Guberina}, it has been shown, and it is
generally accepted, that in the case of a flat
 Universe the most suitable ansatz is the use of the event horizon $R_h$
 \cite{Hsu:2004ri}:
 \begin{equation}
  R_h= a\int_t^\infty \frac{dt}{a}=a\int_a^\infty\frac{da}{Ha^2},
  \label{Rh}
 \end{equation} which leads to results compatible with
 observations.
 The
holographic energy density $\rho_{\Lambda}$ is given by
\begin{equation} \rho_{\Lambda}=\frac{3c^2}{R_{h}^2}
\label{holo},\end{equation} in units where $M_p^2=8\pi$, and $c$
is a constant which value is determined by observational fit.
Furthermore, we can define the dimensionless dark energy as:
\begin{equation}
\Omega_{\Lambda}\equiv\frac{\rho_{\Lambda}}{3H^2}=\frac{c^2}{R_{h}^2H^2}\label{omega}.
\end{equation}
In the case of a dark-energy dominated universe, dark energy
evolves according to the conservation law
\begin{equation}
\dot{\rho}_{\Lambda}+3H(\rho_{\Lambda}+p_{\Lambda})=0
\label{coneq},
\end{equation}
 or equivalently:
\begin{equation}
\label{Omevol}
\dot{\Omega}_{\Lambda}=H\Omega_{\Lambda}(1-\Omega_{\Lambda})
\left(1+\frac{2\sqrt{\Omega_{\rm
\Lambda}}}{c}\right),
\end{equation}
where the equation of state is
\begin{equation}
p_{\Lambda}=-\frac{1}{3}\left(1+\frac{2\sqrt{\Omega_{\Lambda}}}{c}\right)\rho_{\Lambda}
\label{eqstat},
\end{equation}
 which leads straightforwardly to an
index of the equation of state of the form:
\begin{equation}w_{\Lambda}=-\frac{1}{3}\left(1+\frac{2\sqrt{\Omega_{\Lambda}}}{c}\right).
\label{index}
 \end{equation}
As we can clearly see, $w_{\Lambda}$ depends on the
 parameter $c$. In recent fit studies,
different groups have ascribed different values to $c$. A direct
fit of the present available SNe Ia data indicates that the best
fit result is the best-fit value $c=0.21$  within 1-$\sigma$ error
range \cite{HG}. In addition, observational data from the X-ray
gas mass fraction of galaxy clusters lead to $c=0.61$ within
1-$\sigma$ \cite{observHDExray}. Similarly, combining data from
type Ia supernovae, cosmic microwave background radiation and
large scale structure give the best-fit value $c=0.91$ within
1-$\sigma$ \cite{observHDECMB}, while combining data from type Ia
supernovae, X-ray gas and baryon acoustic oscillation lead to
$c=0.73$ as a best-fit value within 1-$\sigma$ \cite{observHDE}.
Finally, the study of the constraints on the dark energy arising
from the holographic connection to the small $l$ CMB suppression,
reveals that $c=2.1$ within 1-$\sigma$ error \cite{cmb3}. In
conclusion, $0.21\leq c\leq 2.1$, and holographic dark energy
provides the mechanism for the $w=-1$ crossing and the transition
to the accelerating expansion of the Universe.

\section{Correspondence between Holographic and
Gauss-Bonnet Dark Energy models}

The main goal of this work is to investigate the conditions under
which there is a correspondence between the Gauss-Bonnet dark
energy model and the holographic dark energy scenario, in the flat
Universe case. In particular, to determine an appropriate
Gauss-Bonnet potential which makes the two pictures to coincide
with each other.

Let us first consider the simple Gauss-Bonnet solutions acquired
in \cite{4,Nojiri2}. In this case $f(\phi)$ is given as \cite{3}
\begin{equation}
\label{fphi} f(\phi)=f_0e^{\frac{2\phi}{\phi_{0}}}.
\end{equation}
In addition, we assume that the scale factor behaves as
$a=a_0t^{h_0}$, and similarly to \cite{4} we will examine both
$h_0$-sign cases. However, when $h_0$ is negative the scale factor
does not correspond to expanding universe but to shrinking one. If
one changes the direction of time as $t\rightarrow -t$, the
expanding universe whose scale factor is given by
$a=a_0(-t)^{h_0}$ emerges. since $h_0$ is not an integer in
general, there is one remaining difficulty concerning the sign of
$t$. To avoid the apparent inconsistency, we may further shift the
origin of the time as $t\rightarrow -t\rightarrow t_s-t$. Then the
time $t$ can be positive as long as $t < t_s$, and we can
consistently write $a=a_0(t_s-t)^{h_0}$. Thus, we can finally
write \cite{4}
\begin{equation}
\label{h0}
 H=\frac{h_0}{t},
\hspace{1cm}\phi=\phi_{0} \ln \frac{t}{t_1}
 \end{equation}
 when
$h_0> 0$ or
\begin{equation}
\label{h01} H=\frac{-h_0}{t_s-t}, \hspace{1cm}\phi=\phi_{0}
\ln \frac{t_s-t}{t_1}
\end{equation}
when $h_0< 0$, with an
undetermined constant $t_1$.

Let us first investigate the positive-$h_0$ case. If we establish
a correspondence between the holographic dark energy and
Gauss-Bonnet approach, then using dark energy density equation
(\ref{1}) and relation (\ref{omega}), together with expressions
(\ref{h0}), we can easily derive the scalar potential term as
\begin{equation}
\label{Vphi}
V=\frac{e^{-\frac{2\phi}{\phi_{0}}}}{t_1^{2}}\left(3\Omega_{\Lambda}h_{0}^{2}+\frac{48f_0h_{0}^{3}}{t_1^{2}}
-\frac{\gamma \phi_0^{2}}{2}\right).
 \end{equation}
Note that expressions (\ref{h0}) allow for an elimination of time
$t$ in terms of the scalar field $\phi$. Furthermore, by
substituting $\phi$, and $H$ from (\ref{h0}), $f(\phi)$ from
(\ref{fphi}) and $V(\phi)$ from (\ref{Vphi}) into (\ref{3}) we
obtain:
\begin{equation}
\label{7}
 -3\gamma
h_0\phi_{0}+\frac{6\Omega_{\Lambda}h_0^{2}}{\phi_{0}}+\frac{96f_0h_0^{3}}{\phi_{0}
t_1^{2}}-3h_0^{2}\frac{d\Omega_{\Lambda}}{d\phi}+\frac{48f_0h_0^{3}(h_0-1)}{\phi_{0}
t_1^{2}}=0
\end{equation}
where
\begin{equation}\frac{d\Omega_{\Lambda}}{d\phi}=\frac{d\Omega_{\Lambda}}{dt}\frac{t}{\phi_{0}}=\frac{d\Omega_{\Lambda}}
{dt}\frac{t_1}{\phi_{0}}\,e^{\frac{\phi}{\phi_{0}}}, \label{8}
\end{equation}
with $\dot{\Omega}_{\Lambda}$ given by (\ref{Omevol}).

Now, under the ansatz $a=a_0t^{h_0}$ it is easy to see from
(\ref{Rh}) that in order for $R_h$ to be finite, $h_0$ has to be
greater than 1. In such a case we straightforwardly find:
\begin{equation}
  R_h=\frac{t}{h_0-1},
\end{equation}
\begin{equation}
\Omega_\Lambda=\frac{c^2(h_0-1)^2}{h_0^2},
\end{equation}
and
\begin{equation}
w_\Lambda=\frac{2}{3h_0}-1.\label{wh0}
\end{equation}
Lastly, in order to obtain a complete consistency in equations
(\ref{7})-(\ref{wh0}), we have to impose the constraint:
\begin{equation}
h_0=\frac{c}{c-1},
\end{equation}
for $c\neq1$, which is moreover consistent with $h_0>1$ (when
$c=1$ we need $h_0=1/2$ which has been excluded due to $R_h$
convergence, i.e in this case there is no accepted solution). Note
however that when $c\neq1$ we must have $c>1$ in order to
``remain'' in the positive-$h_0$ case. As we observe, the case
under examination leads to a correspondence between Gauss-Bonnet
and Holographic dark energy models, where $\Omega_\Lambda=1$, and
$-\frac{1}{3}\geq w_\Lambda\geq-1$ (as it is implied by
(\ref{wh0}) under $h_0>1$). Thus, it is not too practical since it
corresponds to a Universe with complete dominance of dark energy,
and which is not accelerating.

Let us proceed to the investigation of the negative-$h_0$ case.
Repeating the same steps, but imposing relations (\ref{h01}) we
find that
\begin{equation}
\label{Vphi1}
V=\frac{e^{-\frac{2\phi}{\phi_{0}}}}{t_1^{2}}\left(3\Omega_{\Lambda}h_{0}^{2}-\frac{48f_0h_{0}^{3}}{t_1^{2}}
-\frac{\gamma \phi_0^{2}}{2}\right),
 \end{equation}
 and
\begin{equation}
\label{conb} -2\gamma \phi_{0} +3\gamma
h_0\phi_{0}+\frac{6\Omega_{\Lambda}h_0^{2}}{\phi_{0}}+\frac{96f_0h_0^{3}}{\phi_{0}
t_1^{2}}-3h_0^{2}\frac{d\Omega_{\Lambda}}{d\phi}+\frac{48f_0h_0^{3}(h_0-1)}{\phi_{0}
t_1^{2}}=0,
\end{equation}
 where
\begin{equation}\frac{d\Omega_{\Lambda}}{d\phi}=-\frac{d\Omega_{\Lambda}}{dt}\frac{(t_s-t)}{\phi_{0}}=-\frac{d\Omega_{\Lambda}}
{dt}\frac{t_1}{\phi_{0}}\,e^{\frac{\phi}{\phi_{0}}}. \label{81}
\end{equation}

Now, under the ansatz $a=a_0(t_s-t)^{h_0}$ we can see from
(\ref{Rh}) that $R_h$ is always finite if $h_0$ is negative, which
is just the case under investigation. Then we have:
\begin{equation}
  R_h=\frac{t_s-t}{1-h_0},
\end{equation}
\begin{equation}
\Omega_\Lambda=\frac{c^2(h_0-1)^2}{h_0^2},
\end{equation}
and therefore
\begin{equation}
w_\Lambda=\frac{2}{3h_0}-1.\label{wh01}
\end{equation}
Finally, in order to obtain a complete consistency in equations
(\ref{conb})-(\ref{wh01}), we have to impose the constraint:
\begin{equation}
h_0=\frac{c}{c-1},
\end{equation}
while there is no accepted solution if $c=1$. Furthermore, since
we are in the negative-$h_0$ case we must have $c<1$. Thus, we
conclude that we acquire a correspondence between Gauss-Bonnet and
Holographic dark energy models. In addition, it is interesting
that in this case we get $w_\Lambda\leq-1$ which corresponds to an
accelerating universe.

As we can see so far, under the ansatz $a=a_0t^{h_0}$ (for
$h_0>0$) or $a=a_0(t_s-t)^{h_0}$ (for $h_0<0$), which is usually
assumed in the Gauss-Bonnet models of dark energy
\cite{4,Nojiri2}, we do obtain a correspondence with holographic
dark energy scenario. Although this correspondence might look
rather trivial, we do acquire an accelerating universe and
moreover an interesting classification in terms of the parameter
$c$ of holographic dark energy, since $c>1$ corresponds to $h_0>1$
and $c<1$ to $h_0<0$.

Let us now make some comments about the freedom of construction of
Gauss-Bonnet or holographic dark energy models separately. In the
basic work \cite{Nojiri2}, as well as in \cite{Nojiri3}, the
authors examine several examples of scalar-Gauss-Bonnet gravity,
with multiple fields (phantom or canonical). Starting from various
ansatzes  for the form of $f(\phi)$, they study the cosmological
evolution. However, simple algebraic calculations reveal that
these models, although correct from the Gauss-Bonnet dark energy
point of view, are not consistent with the holographic dark energy
framework. Specifically, we find that the evolution for
$\Omega_\Lambda$ is not consistent with (\ref{Omevol}). Similarly,
starting with conventional holographic dark energy models it is
not a priori ensured that these models fit within the Gauss-Bonnet
framework. Therefore, one must be careful in constructing either
of the two scenarios. In particular, he must ensure the
simultaneous validity of relations (\ref{3}), (\ref{1}) and
(\ref{Omevol}). Thus, these relations correspond to specific
constraints for the function $f(\phi)$, which describes the
coupling with gravity, and for the potential $V(\phi)$.

An enlightening contribution to the aforementioned discussion is
the following. The peculiar feature of holographic dark energy, as
long as one accepts its (still controversial in the literature)
framework, is that it correlates dark energy with the event
horizon of the universe, i.e. with the scale factor (or
equivalently with the Hubble parameter). Thus, apart from the
Friedmann equations, and the evolution equations for the scalar
fields, one has to fulfill this additional correlation, which is
quantitatively expressed by relation (\ref{omega}) (or
(\ref{Omevol})). In other words, this relation can be considered
as an extra constraint, imposed externally to the system of
Einstein equations. On the other hand, in Gauss-Bonnet framework,
dark energy acquires a contribution from the scalar field and its
potential, and from the Gauss-Bonnet coupling, namely relation
(\ref{1}). In \cite{Nojiri2}, the authors reconstruct the
scalar-Gauss-Bonnet gravity for general cosmological evolutions,
i.e they reconstruct the scalar field potential $V(\phi)$ and the
coupling $f(\phi)$ for arbitrary $H(t)$ and $\phi(t)$. Although
this procedure is efficient for any conventional cosmological
evolution, it is not adequate for the peculiar and
non-conventional holographic dark energy evolution, due to the
external, additional and non-trivial constraint. In particular, in
order to satisfy relation (\ref{1}) in full generality, and for an
arbitrary $\rho_\Lambda$ (which will be later identified with
relation (\ref{Omevol})), one has to specify $H(t)$, $f(\phi)$,
$\phi(t)$ and $V(\phi)$ independently, which cannot be performed
using the procedure of \cite{Nojiri2}. The peculiar nature of
holographic dark energy framework can be embedded in Gauss-Bonnet
gravity only by the simultaneous satisfaction of relations
(\ref{3}), (\ref{1}) and (\ref{Omevol}).

\section{Conclusions}

In the present paper, we  considered separately the holographic
and Gauss-Bonnet dark energy models. Then we investigated the
conditions under which the two scenarios can be simultaneously
valid. In particular, by considering holographic dark energy
density as a dynamical cosmological constant, we obtained its
equation of state in the Gauss-Bonnet framework. Thus, we have
suggested  a correspondence between the holographic dark energy
scenario in flat universe and the phantom dark energy model in the
framework of Gauss-Bonnet theory with a potential. This
correspondence can lead to an accelerating universe, under a
special ansatz for the function $f(\phi)$, which describes the
coupling with gravity, and for the scale factor evolution.
However, in general one has not full freedom of constructing
independently the two cosmological scenarios. If he requires their
consistent unification, then specific constraints must be imposed
in  $f(\phi)$, as well as in the potential $V(\phi)$.

\end{document}